\newcommand{\PRE}[1]{}
\documentclass[prl,twocolumn,amsmath,preprintnumbers]{revtex4}

\usepackage{epsfig}

\newcommand{\be}{\begin{equation}}
\newcommand{\ee}{\end{equation}}
\newcommand{\bea}{\begin{eqnarray}}
\newcommand{\eea}{\end{eqnarray}}

\newcommand{\ba}{\begin{array}}
\newcommand{\ea}{\end{array}}

\newcommand{\lsim}{
\mathrel{\hbox{\rlap{\hbox{\lower4pt\hbox{$\sim$}}}\hbox{$<$}}}}
\newcommand{\gsim}{
\mathrel{\hbox{\rlap{\hbox{\lower4pt\hbox{$\sim$}}}\hbox{$>$}}}}

\newcommand{\tev}{\text{TeV}}
\newcommand{\gev}{\text{GeV}}

\newcommand{\fb}{\text{fb}}

\begin{document}

\preprint{UCRHEP-T483}
\title{
\PRE{\vspace*{1.5in}}
Gauged $B-x_iL$ origin of $R$ parity and its implications
\PRE{\vspace*{0.3in}} }
\author{Hye-Sung Lee}
\affiliation{Department of Physics, Brookhaven National Laboratory, 
Upton, NY 11973, USA}
\author{Ernest Ma}
\affiliation{Department of Physics and Astronomy, University of California, 
Riverside, CA 92521, USA}
\date{January, 2010}

\begin{abstract}
\PRE{\vspace*{.1in}} \noindent
Gauged $B-L$ is a popular candidate for the origin of the conservation of 
$R$ parity, i.e. $R = (-)^{3B+L+2j}$, in supersymmetry, but it fails to forbid 
the effective dimension-five terms arising from the superfield combinations 
$QQQL$, $u^cu^cd^ce^c$, and $u^cd^cd^cN^c$, which allow the proton to decay. 
Changing it to $B-x_iL$, where $x_e+x_\mu+x_\tau=3$ (with $x_i \neq 1$) for the 
three families, would forbid these terms while still serving as a gauge 
origin of $R$ parity.  We show how this is achieved in 
two minimal models with realistic neutrino mass matrices, and discuss 
their phenomenological implications.
\end{abstract}

\pacs{11.30.Er, 12.60.-i, 14.70.Pw}

\maketitle

\subsubsection{Introduction}
In the Minimal Supersymmetric Standard Model (MSSM) of particle interactions, 
the imposition of $R$ parity, i.e. $R = (-)^{3B+L+2j}$, where $B$, $L$, and $j$ stand for baryon number, lepton number, and spin, respectively, serves two purposes. 
One is to establish a candidate for the dark-matter of the Universe, because 
the lightest supersymmetric particle (LSP) is absolutely stable, being odd 
under $R$. The other is to forbid the otherwise allowed renormalizable superfield terms $L H_u$, $L L e^c$, $L Q d^c$, and $u^c d^c d^c$, so that 
the proton does not decay as a result of these interactions.  However, the 
higher-dimensional quadrilinear terms $Q Q Q L$ and $u^c u^c d^c e^c$ are 
still allowed, giving rise thus to effective dimension-five terms in the 
Lagrangian which would also induce fast proton decay \cite{Weinberg:1981wj}.  
This is a serious problem for grand unification, even at a scale as high 
as $10^{16}$ GeV.

With the addition of three neutral singlet superfields $N^c$, desirable for 
neutrino masses, gauged $B-L$ becomes possible.  In that case, $R$ parity 
is better understood theoretically as a discrete remnant of $B-L$ breaking.  
However, the offending quadrilinear terms remain, together with the new term 
$u^c d^c d^c N^c$.  To get rid of these terms, we propose a simple solution.  
Instead of gauged $B-L$, we adopt a flavor-dependent variation, i.e. 
$B - x_i L$, where the usual leptons still have $L=1$, but $x_{e,\mu,\tau}$ 
for the three families are not equal to 1.  In fact, the idea that there is 
just one $N^c$ and that $x_{e,\mu,\tau} = (3,0,0)$ or $(0,3,0)$ or $(0,0,3)$ 
has already been explored in the past~\cite{m98,ms98,mr98,bhmz09,Salvioni:2009jp}.
If $x_i \neq 1$, then all the unwanted quadrilinear terms are forbidden by 
gauged $B - x_i L$ so that proton stability is assured as shown for 
$(0,0,3)$ case in Ref.~\cite{ms98}.

Here we assume that there are three $N^c$, with each $x_i$ nonzero.
Unlike the previous $(0,0,3)$ type of models, this opens up a new possibility
that the spontaneous breaking of $B - x_i L$ by suitably chosen Higgs singlet 
superfields $S_{1,2}$ 
will result in an exact discrete residual symmetry which is 
just the usual $R$ parity. In the following, we will discuss the conditions 
on $x_i$ and construct two minimal models with realistic neutrino mass 
matrices.  We will examine their phenomenological constraints and the 
prognosis of their verification at the Large Hadron Collider (LHC).

\subsubsection{Model}
Our minimal model consists of the usual particle content of the MSSM plus 
three singlet superfields $N^c_i$ with $L=-1$ and two singlet superfields 
$S_{1,2}$ with $L = \pm 2$.  Under the gauged $U(1)_X$ symmetry of $B-x_iL$, 
quarks have charges 1/3 and leptons have charges $-x_i$.  The two usual 
Higgs doublet superfields have charges 0 and the Higgs singlets $S_{1,2}$ 
have charges $\mp 2x_S$.
The various $x_i$ and $x_S$ will be determined by the
requirements of anomaly cancellation and a realistic neutrino mass matrix.

The anomaly-free conditions for the addition of $U(1)_X$ to the standard 
$SU(3)_C \times SU(2)_L \times U(1)_Y$ gauge group are easily written down. 
The $[SU(3)_C]^2 U(1)_X$ and $[U(1)_X]^3$ anomalies are automatically zero 
because of the vectorial nature of $SU(3)_C$ and $U(1)_X$.
The $[\text{gravity}]^2 U(1)_X$ anomaly also vanishes for the same reason.
The $[U(1)_X]^2 U(1)_Y$ anomaly is zero because the sum of $Y$ charges is 
zero separately for each family of quarks and for each family of leptons. 
The remaining two conditions, i.e. $[SU(2)_L]^2 U(1)_X$ and 
$[U(1)_Y]^2 U(1)_X$, are given respectively by
\begin{eqnarray}
&& (3)(3)(1/3) - x_e - x_\mu - x_\tau = 0,\\ 
&& (3)(3)[2(1/6)^2-(2/3)^2-(-1/3)^2](1/3)\nonumber \\
&& +~[2(-1/2)^2-(-1)^2](-x_e-x_\mu-x_\tau) = 0.
\end{eqnarray}
Both are satisfied if
\begin{equation}
x_e + x_\mu + x_\tau = 3. \label{eq:sumrule}
\end{equation}
The usual $B-L$ is recovered for $x_e=x_\mu=x_\tau=1$, and 
$B-3L_\tau$~\cite{m98} is obtained for $x_e=x_\mu=0$ and $x_\tau=3$.

As $S_{1,2}$ acquire vacuum expectation values, $U(1)_X$ is broken.
$(x_e, x_\mu, x_\tau)$ as well as $x_S$ will be chosen in such 
a way that a residual symmetry of $B-x_iL$ will remain which is exactly 
$R$ parity, and that realistic neutrino masses and mixing are obtained 
via the canonical seesaw mechanism.  We note that since the Higgs doublets 
do not transform under $U(1)_X$ and $S_{1,2}$ do not transform under 
$SU(3)_C \times SU(2)_L \times U(1)_Y$, the resulting $Z$ and $Z'_X$ bosons 
do not mix. This avoids the stringent constraint from precision electroweak 
measurements at the $Z$ resonance. (See Ref.~\cite{Erler:2009jh} and 
references therein.)

To obtain $R$ parity (equivalently, matter parity) as a {\em total} 
residual symmetry from the breaking of $B-x_iL$ using 
$S_{1,2}$, these new singlet superfields (with $L = \pm 2$ and $B=0$) should 
satisfy $|x_S| = 1/3$.
Since $B=1/3$ for quarks and $L=1$ for leptons, we obtain the usual definition 
of $R$ parity for all particles if $3 x_{e,\mu,\tau}$ are odd integers.  
For a detailed discussion about conditions to get a $Z_N$ out of a $U(1)$ 
gauge symmetry, see Ref.~\cite{hll09} and references therein.

Flavor-dependent $U(1)$ models have been widely studied to address many 
issues (e.g., see Ref.~\cite{Langacker:2000ju}).
Our idea of having a particular discrete symmetry out of a flavor-dependent 
$U(1)$ gauge symmetry can be viewed as a useful guide in constraining 
such models.

\subsubsection{Neutrino sector}
The general requirements discussed above would still allow an infinite 
number of possible models, until realistic neutrino masses and mixing are 
considered.

If all $x_i$ are different for the three families of leptons, the 
charged lepton mass matrix and the Dirac mass matrix linking $\nu_i$ with 
$N^c_j$ are both constrained to be diagonal.  To obtain mixing among the 
three neutrinos, the $3 \times 3$ Majorana matrix spanning $N^c_j$ must 
have enough nonzero entries.  However, the only sources of such terms 
are $S_{1,2} N_j^c N_k^c$ and $N_j^c N_k^c$ if $x_j+x_k = 0$.  With three 
different $x_{e,\mu,\tau}$ and just $\pm x_S$ to work with, this is clearly 
impossible.

We now assume that $x_\mu = x_\tau$, then some simple algebra will show 
that there are only two solutions (recalling that $S_{1,2}$ have $L = \pm 2$):
\begin{eqnarray}
\text{(I)} && x_S = -x_e = x_\mu, \\
\text{(II)} && x_S = -x_\mu = (x_e + x_\mu)/2.
\end{eqnarray}
Together with Eq.~\eqref{eq:sumrule}, this means that
\begin{eqnarray}
\text{(I)} && x_{e,\mu,\tau} = (-3,3,3),~~ x_S = 3, \\
\text{(II)} && x_{e,\mu,\tau} = (9,-3,-3),~~ x_S = 3.
\end{eqnarray}
In model (I), the $3 \times 3$ Majorana mass matrix spanning $N^c_j$ has all 
nonzero entries: $N_e^c N^c_{\mu,\tau}$ are invariant mass terms, $N^c_e N^c_e$ 
comes from $\langle S_2 \rangle$, and $N^c_{\mu,\tau} N^c_{\mu,\tau}$ 
come from $\langle S_1 \rangle$.  In model (II), the $N^c_e N^c_e$ entry 
is zero, but all others are nonzero: $N^c_{\mu,\tau} N^c_{\mu,\tau}$ come from 
$\langle S_2 \rangle$, and $N_e^c N^c_{\mu,\tau}$ from 
$\langle S_1 \rangle$.  Both are general enough for obtaining a 
realistic neutrino mass matrix, with mixing among all three lepton families.

\subsubsection{Baryon triality}
As discussed above, the requirement of a realistic neutrino mass matrix 
using only $S_{1,2}$ demands $|x_S| = 3$ instead of $|x_S| = 1/3$.
This means that the {\em total} discrete symmetry from $B-x_i L$ is not 
just $R$ parity any more. It has been extended to a larger symmetry.
Following the general arguments in Ref.~\cite{hll09}, we find that our 
total discrete symmetry is now $Z_6$, which is a direct product of $R$ 
parity and baryon triality ($Z_6 = R_2 \times B_3$).
Under baryon triality ($B_3 = Z_3$)~\cite{Ibanez:1991pr}, baryons transform 
as $\omega = \exp (2 \pi i/3)$, so that the proton is absolutely 
stable, being the lightest particle with that charge.  This result came 
as a pleasant surprise, because it was not our intention to 
construct a model which contains $B_3$.  It points to a possible deep connection between neutrino mass and proton stability.

In regard to baryon stability, there are other relevant anomaly-free $U(1)$ gauge symmetry models~\cite{hll09,Babu:2003qh,l08}, as well as discrete gauge symmetry models~\cite{Ibanez:1991pr,Babu:2002tx,dlt06}.
Especially, we note that when model (I) is shifted by some hypercharge, it can reach the equivalent form of a model in Ref. \cite{Babu:2003qh}.

\subsubsection{$e-\mu-\tau$ nonuniversality}
The salient prediction of our proposal is the existence of a new neutral 
gauge boson $Z'_X$.  It does not mix with the electroweak $Z$ at tree level, 
but it couples 
to all quarks and leptons in a specified way.  In particular, it breaks 
$e-\mu-\tau$ universality.  Thus it may be important as a one-loop effect 
\cite{mr98} in the precision measurements of $Z \to \ell^+ \ell^-$.  However, 
this effect is proportional to $x_i^2$, and its contribution to nonuniversality 
is zero for model (I).  As for 
model (II), the prediction is that $\Gamma(Z \to e^+e^-)$ should be bigger 
than $\Gamma(Z \to \mu^+ \mu^-)$ and $\Gamma(Z \to \tau^+ \tau^-)$, 
and it is proportional to $(81-9)g_X^2$.  The present world averages 
are~\cite{pdg08} 
\begin{eqnarray}
\Gamma_e &=& 83.91 \pm 0.12~\text{MeV},\\
\Gamma_\mu &=& 83.99 \pm 0.18~\text{MeV},\\
\Gamma_\tau &=& 84.08 \pm 0.22~\text{MeV}.
\end{eqnarray}
After adding a kinematical correction of 0.19 MeV to $\Gamma_\tau$, we 
find the deviation of $\Gamma_e$ from the average of $\Gamma_\mu$ and 
$\Gamma_\tau$ to be bounded at 95\% C.L. by
\begin{equation}
\Delta \Gamma_e/\Gamma_{\mu,\tau} < 0.002.
\end{equation}
Let $r \equiv M_{Z'_X}^2/M_Z^2$, then the one-loop radiative correction to 
$Z \to \ell^+ \ell^-$ from $Z'_X$ exchange in model (II) is given by \cite{cm95}
\begin{equation}
{\Delta \Gamma_e \over \Gamma_{\mu,\tau}} = {(81-9) g_X^2 \over 8 \pi^2} F_2(r),
\end{equation}
where
\begin{eqnarray}
&& F_2(r) = -{7 \over 2} - 2r - (2r + 3) \ln r + 2(1+r)^2 \times \nonumber \\ 
&& \left[ {\pi^2 \over 6} - \text{Li}_2 \left( {r \over 1+r} \right) 
- {1 \over 2} \ln^2 \left( {r \over 1+r} \right) \right].
\end{eqnarray}
In the above, $\text{Li}_2(x) = -\int^x_0 (dt/t)\ln(1-t)$ is the Spence 
function. For $r \gg 1$ (i.e. $M^2_{Z'_X} \gg M^2_Z$) which will be required by 
Tevatron data in any case (as shown below), $F_2 \simeq r^{-1} [11/9 + 
(2/3)\ln r]$ and the resulting numerical bound is
\begin{equation}
g_X^2/M^2_{Z'_X} \lsim 0.05~\tev^{-2}, \label{eq:bound}
\end{equation}
to a good approximation.  There is also an overall correction to 
$\Gamma(Z \to \text{hadrons})/\Gamma(Z \to \text{leptons})$, but 
that effect may be absorbed into the value of $\alpha_S$ in QCD.

\subsubsection{LEP2 contact interaction}
The agreement of the SM and the LEP2 data provides the indirect bounds on the $Z'_X$ mass.
The strongest one for our models (I) and (II), where $\mu$ and $\tau$ have the same charges and $e$ has the charge of opposite sign, comes from the $e^+e^- \to \mu^+\mu^-$, $\tau^+\tau^-$ channel with $\Lambda_{VV}^- = 16 ~\tev$ \cite{:2003ih}. (See Ref.~\cite{Carena:2004xs} for some useful discussions.)

The $Z'_X$ mass is bounded by
\be
M_{Z'_X}^2 \gsim {g_X^2 \over 4\pi} |x_e x_\mu| (\Lambda_{VV}^-)^2
\ee
for sufficiently large $M_{Z'}$ compared to the LEP2 energy.
It results
\bea
M_{Z'_X} &\gsim& 1.4 ~\tev \quad \text{in model (I)}, \\
M_{Z'_X} &\gsim& 2.3 ~\tev \quad \text{in model (II)},
\eea
for $g_X = 0.1$.

\subsubsection{Phenomenology of $Z'_X$}
The direct production of $Z'_X$ is possible at hadron colliders through 
its coupling to quarks.  Its decay to leptons is then a clear 
signature. At present, there is a Tevatron limit on the cross section 
$\sigma(p \bar{p} \to Z'_X \to e^+ e^-)$ at a center-of-mass energy 
$E_\text{cm} = 1.96$ TeV, based on an integrated luminosity of $L = 2.5$ 
fb$^{-1}$ \cite{cdf_e}, and similarly for dimuons \cite{cdf_mu}.  To 
compare against these results, we take $g_{X} = 0.1$ for definiteness. 
We assume $\Gamma_{Z'_X} = 0.01 M_{Z'_X}$ for model (I) and $\Gamma_{Z'_X} = 
0.04 M_{Z'_X}$ for model (II), which are approximately their true values if  
they only decay into particles of the Standard Model (SM), i.e. not their 
superpartners nor the additional singlets.  

\begin{figure}[tb]
\begin{center}
\includegraphics[width=0.4\textwidth]{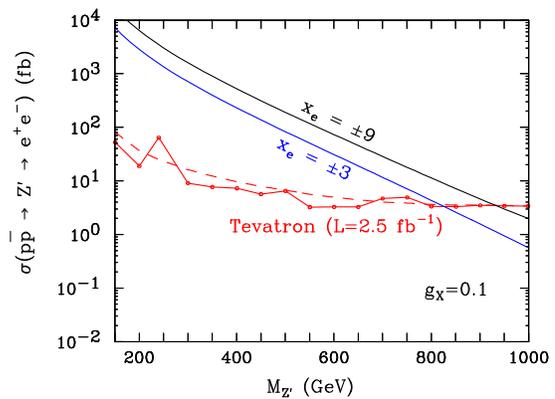}
\end{center}
\caption{
Tevatron bounds on $Z'_X$ mass in models (I) and (II).
}
\label{fig:Tevatron}
\end{figure}

For the numerical analysis, we use {\tt CompHEP/CalcHEP} 
\cite{Pukhov:1999gg,Pukhov:2004ca} and the parton distribution functions 
of {\tt CTEQ6L} \cite{Pumplin:2002vw}.  Since the Tevatron bounds from 
dielectron and dimuon data are similar for the same coupling, we show only 
the dielectron result.  We find $M_{Z'_X} \gsim 830 ~\gev$ for model (I) and 
$M_{Z'_X} \gsim 940 ~\gev$ for model (II) as shown in Fig.~\ref{fig:Tevatron}.
(Since $|x_\mu| < |x_e|$ in model (II), the bound from the dimuon data is less 
stringent.) We note also that the $e-\mu-\tau$ nonuniversality constraint 
of Eq.~\eqref{eq:bound} is easily satisfied.

For the LHC discovery reach (with the design energy $E_\text{cm} = 14 ~\tev$) 
through the 
dilepton $Z'_X$ resonance, we use cuts $p_T > 20 ~\gev$, $|\eta| < 2.4$ 
(for each lepton) and $|m_\text{inv}(\ell^+ \ell^-)| < 3 \Gamma_{Z'_X}$.
SM background at the LHC with these cuts is negligible, and we just require 
10 signal events to claim its discovery at the LHC for a fixed flavor.

\begin{figure*}[tb]
\begin{center}
\includegraphics[width=0.4\textwidth]{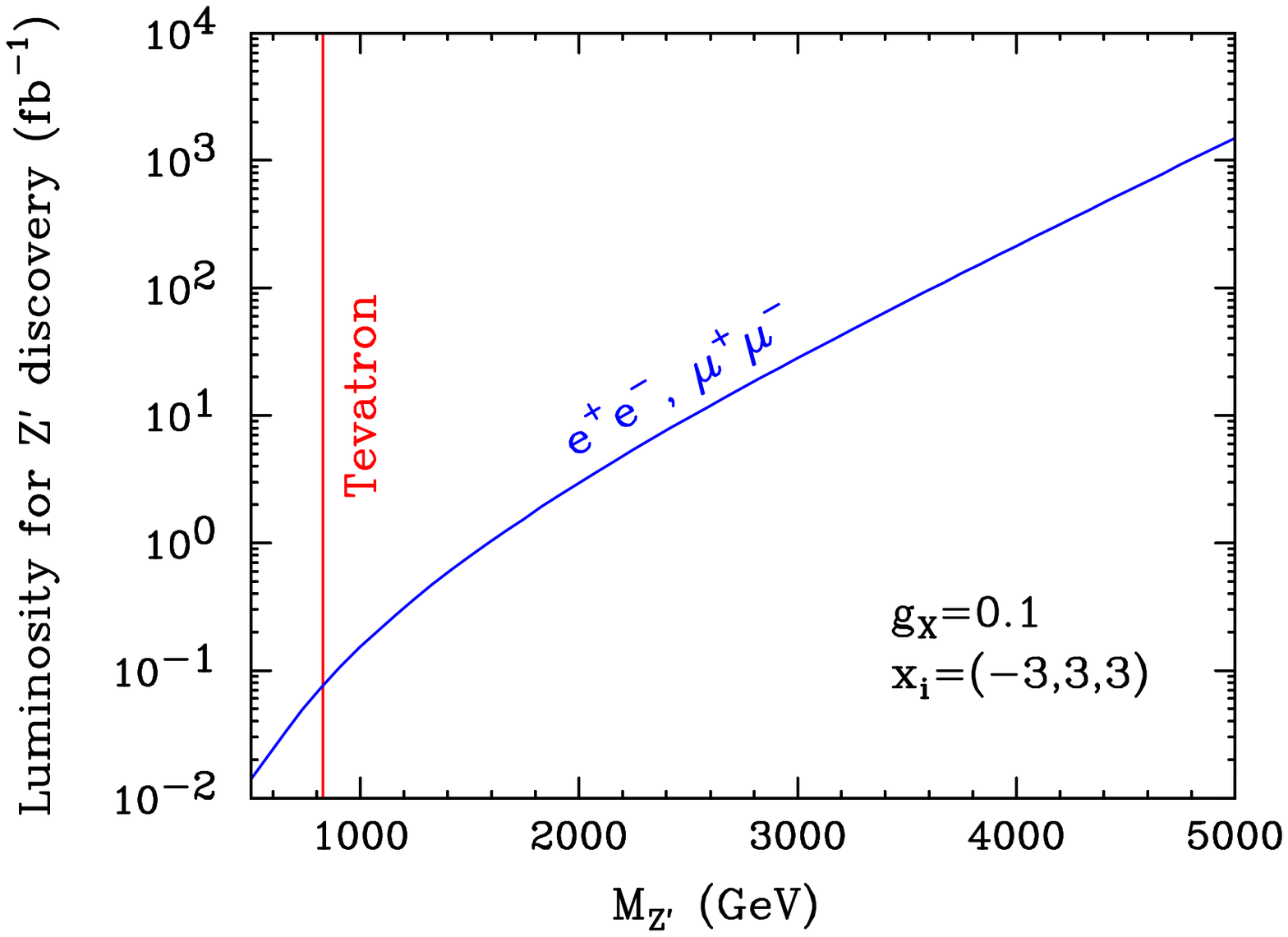} ~~~
\includegraphics[width=0.4\textwidth]{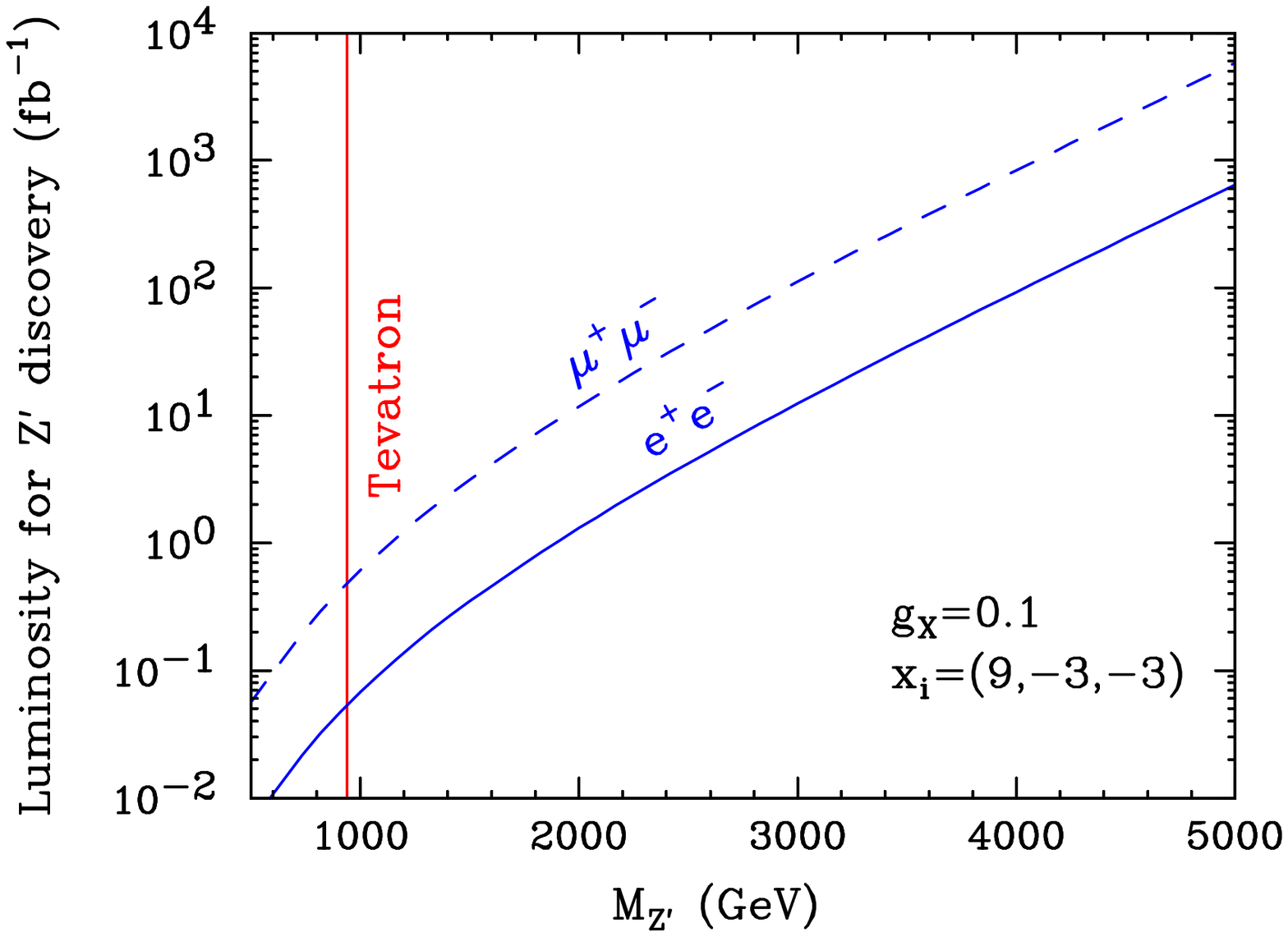} \\
(a) ~~~~~~~~~~~~~~~~~~~~~~~~~~~~~~~~~~~~~~~~~~~~~~~~~~~~~~~~~~~~~~~~~ (b)
\end{center}
\caption{
The LHC discovery reach for (a) model (I) and (b) model (II). 
The required luminosity is the same for both $e^+e^-$ and $\mu^+\mu^-$ 
resonances in model (I). 
The required luminosity for the $e^+e^-$ (solid line) resonance is about an 
order of magnitude smaller than that for the $\mu^+\mu^-$ (dashed line) 
resonance in model (II).
}
\label{fig:LHC}
\end{figure*}

Figs.~\ref{fig:LHC}(a) and 2(b) show the LHC discovery reach using 
dileptons for models (I) and (II) respectively.  From the dilepton resonance 
only, model (I) cannot be distinguished from the flavor-independent case of 
$B-L$.  In model (I), $Z'_X$ will be revealed by both $e^+e^-$ and 
$\mu^+\mu^-$ channels at the same luminosity ($L \simeq 1 ~\fb^{-1}$ for 
$M_{Z'_X} = 1.5 ~\tev$).  In model (II), the $\mu^+\mu^-$ resonance will need 
a luminosity about an order of magnitude larger than that for the $e^+e^-$ 
resonance because $\sigma(\mu^+\mu^-)/\sigma(e^+e^-) \simeq (-3)^2/(9)^2$.

Since Higgs doublets have zero charges under $B - x_i L$, the channels which 
require nonzero charges such as the 6-lepton resonance discussed in 
Ref.~\cite{Barger:2009xg} will be absent.

In the presence of $Z'_X$ at the TeV scale, a (predominantly right-handed) 
sneutrino as well as the usual neutralino can be a good LSP dark-matter 
candidate~\cite{Lee:2007mt}.

\subsubsection{One-loop charged lepton processes}
There are two sources of lepton flavor violation.  One comes 
from explicit interactions linking $N^c$ through $S_{1,2}$; the other 
through the mismatch of lepton and slepton mass matrices (which is 
common to all supersymmetric models). Since we choose $x_e \neq x_\mu = x_\tau$, 
the latter applies only to the $\mu-\tau$ sector, whereas the former  
applies to all leptons, but for one-loop processes such as $\mu \to e \gamma$, 
they are negligible because of the smallness of neutrino masses.

The muon anomalous magnetic moment ($a_\mu$) does not violate lepton flavor, 
so it has a contribution from $Z'_X$, i.e.~\cite{jn09}
\begin{equation}
\Delta a_\mu = {3 g_X^2 \over 4 \pi^2} {m_\mu^2 \over M_{Z'_X}^2}
\end{equation}
for both models (I) and (II). Using the Tevatron bounds on $M_{Z'_X}$ with 
$g_X = 0.1$, we find this to be at most of order $10^{-11}$, which is 
negligible compared to the present experimental accuracy.  On the other 
hand, since we are considering a supersymmetric model, there are one-loop 
contributions to $a_\mu$ from the neutralinos and charginos, which may 
account for the possible deviation in its measurement at the Brookhaven 
Alternating Gradient Synchrotron \cite{Bennett}.

\subsubsection{Conclusion}
We have proposed a new $U(1)$ gauge symmetry $B - x_i L$, with 
$x_{e,\mu,\tau} = (-3,3,3)$ [model (I)] or $(9,-3,-3)$ [model (II)], in the 
context of supersymmetry with three neutral singlets $N^c_{e,\mu,\tau}$.  
The spontaneous breaking of this $U(1)_X$ by the addition of singlets 
$S_{1,2}$ with $L = \pm 2$ and $x_S = 3$ accomplishes three objectives. 
(i) The conventional $R$ parity survives as an exact discrete symmetry, 
desirable for having the LSP as a good dark-matter candidate.
(ii) Realistic neutrino masses and mixing are obtained. (iii) An exact 
residual $Z_3$ 
symmetry, i.e. baryon triality $B_3$, emerges which makes the proton 
absolutely stable.

The neutral gauge boson of this new $U(1)_X$ has large couplings to leptons 
over quarks: three or nine times larger than in the case of $B-L$.  Using 
present Tevatron bounds with $g_X=0.1$, we find $M_{Z'_X}$ to be 
greater than $830 ~\gev$ in model (I), and $940 ~\gev$ in model (II).
The indirect bounds from LEP2 are $1.4 ~\tev$ in model (I), and $2.3 ~\tev$ in model (II).
It should be accessible at the LHC, possibly at a very early stage.  
To have a background-free signal of 10 dilepton events at the LHC,  
the $Z'_X$ with $M_{Z'_X} = 1.5 ~\tev$ in model (I) may be discovered for 
an integrated 
luminosity of only about 1 \fb$^{-1}$.  In model (II), because of the 
flavor-dependent charges, the event rate of the $e^+e^-$ resonance is 
predicted to be nine times that of $\mu^+\mu^-$.

\vspace{0.5cm}
This work was supported by U.~S.~Department of Energy Grants 
No. DE-AC02-98CH10886 (HL) and No. DE-FG03-94ER40837 (EM).




\begin{thebibliography}{99}
\bibitem{Weinberg:1981wj}
  S.~Weinberg,
  Phys.\ Rev.\  D {\bf 26}, 287 (1982).
  
\bibitem{m98}
  E.~Ma,
  Phys.\ Lett.\  B {\bf 433}, 74 (1998).
  
\bibitem{ms98}
  E.~Ma and U.~Sarkar,
  Phys.\ Lett.\  B {\bf 439}, 95 (1998).

\bibitem{mr98}
  E.~Ma and D.~P.~Roy,
  Phys.\ Rev.\  D {\bf 58}, 095005 (1998).
  
\bibitem{bhmz09}
  X.~J.~Bi, X.~G.~He, E.~Ma and J.~Zhang,
  arXiv:0910.0771 [hep-ph].

\bibitem{Salvioni:2009jp}
  E.~Salvioni, A.~Strumia, G.~Villadoro and F.~Zwirner,
  arXiv:0911.1450 [hep-ph].

\bibitem{Erler:2009jh}
  J.~Erler, P.~Langacker, S.~Munir and E.~R.~Pena,
  JHEP {\bf 0908}, 017 (2009).

\bibitem{hll09}
  T.~Hur, H.~S.~Lee and C.~Luhn,
  JHEP {\bf 0901}, 081 (2009).

\bibitem{Langacker:2000ju}
  P.~Langacker and M.~Plumacher,
  Phys.\ Rev.\  D {\bf 62}, 013006 (2000).

\bibitem{Ibanez:1991pr}
  L.~E.~Ibanez and G.~G.~Ross,
  Nucl.\ Phys.\  B {\bf 368}, 3 (1992).
  
\bibitem{Babu:2003qh}
  K.~S.~Babu, I.~Gogoladze and K.~Wang,
  Phys.\ Lett.\  B {\bf 570}, 32 (2003).

\bibitem{l08}
  H.~S.~Lee,
  Phys.\ Lett.\  B {\bf 663}, 255 (2008).

\bibitem{Babu:2002tx}
  K.~S.~Babu, I.~Gogoladze and K.~Wang,
  Nucl.\ Phys.\  B {\bf 660}, 322 (2003).
  
\bibitem{dlt06}
  H.~K.~Dreiner, C.~Luhn and M.~Thormeier,
  Phys.\ Rev.\  D {\bf 73}, 075007 (2006).

\bibitem{pdg08}
  [Particle Data Group] C.~Amsler {\it et al.},
  Phys.\ Lett.\  B {\bf 667}, 1 (2008).

\bibitem{cm95}
  C.~D.~Carone and H.~Murayama,
  Phys.\ Rev.\ Lett.\  {\bf 74}, 3122 (1995);
  Phys.\ Rev.\  D {\bf 52}, 484 (1995).

\bibitem{:2003ih}
  [The LEP Collaborations ALEPH, DELPHI, L3, OPAL, the LEP Electroweak Working Group, the SLD Electroweak and Heavy Flavour Groups].
  arXiv:hep-ex/0312023.

\bibitem{Carena:2004xs}
  M.~S.~Carena, A.~Daleo, B.~A.~Dobrescu and T.~M.~P.~Tait,
  Phys.\ Rev.\  D {\bf 70}, 093009 (2004).
      
\bibitem{cdf_e}
  [CDF Collaboration] T.~Aaltonen {\it et al.},
  Phys.\ Rev.\ Lett.\  {\bf 102}, 031801 (2009).

\bibitem{cdf_mu}
  [CDF Collaboration] T.~Aaltonen {\it et al.},
  Phys.\ Rev.\ Lett.\  {\bf 102}, 091805 (2009).

\bibitem{Pukhov:1999gg}
  A.~Pukhov {\it et al.},
  arXiv:hep-ph/9908288.
  
\bibitem{Pukhov:2004ca}
  A.~Pukhov,
  arXiv:hep-ph/0412191.
    
\bibitem{Pumplin:2002vw}
  J.~Pumplin, D.~R.~Stump, J.~Huston, H.~L.~Lai, P.~M.~Nadolsky and W.~K.~Tung,
  JHEP {\bf 0207}, 012 (2002).

\bibitem{Barger:2009xg}
  V.~Barger, P.~Langacker and H.~S.~Lee,
  Phys.\ Rev.\ Lett.\  {\bf 103}, 251802 (2009).

\bibitem{Lee:2007mt}
  H.~S.~Lee, K.~T.~Matchev and S.~Nasri,
  Phys.\ Rev.\  D {\bf 76}, 041302 (2007).
        
\bibitem{jn09}
  F.~Jegerlehner and A.~Nyffeler,
  Phys.\ Rept.\  {\bf 477}, 1 (2009).
    
\bibitem{Bennett}
  [Muon g-2 Collaboration] G.~W.~Bennett {\it et al.},
  Phys.\ Rev.\ Lett.\  {\bf 92}, 161802 (2004);
  Phys.\ Rev.\  D {\bf 73}, 072003 (2006).
  
\end{thebibliography}
\end{document}